\documentclass[acmtog]{acmart}

\AtBeginDocument{\providecommand\BibTeX{{\normalfont B\kern-0.5em{\scshape i\kern-0.25em b}\kern-0.8em\TeX}}}
\acmSubmissionID{1820}

\copyrightyear{2025}
\acmYear{2025}
\setcopyright{cc}
\setcctype{by}
\acmConference[SA Conference Papers '25]{SIGGRAPH Asia 2025 Conference Papers}{December 15--18, 2025}{Hong Kong, Hong Kong}
\acmBooktitle{SIGGRAPH Asia 2025 Conference Papers (SA Conference Papers '25), December 15--18, 2025, Hong Kong, Hong Kong}\acmDOI{10.1145/3757377.3763930}
\acmISBN{979-8-4007-2137-3/2025/12}

\usepackage{multirow}
\usepackage{array}
\usepackage{cleveref}
\usepackage{wrapfig}            % Optional: keep only if it renders fine without overfulls
\usepackage{graphicx}           % (already loaded by acmart, but harmless)
\DeclareGraphicsExtensions{.pdf,.png,.jpg}
\usepackage{makecell}
\usepackage{enumitem}
\usepackage{listings}

\usepackage{savesym}
\savesymbol{zifour@default}
\savesymbol{zifour@scaled}
\usepackage{multirow} 
\usepackage{array}
\usepackage{cleveref}
\usepackage{inconsolata}
\usepackage{spverbatim}
\usepackage{wrapfig}

\usepackage{booktabs}
\usepackage{tabularx}        % auto-sizing tables to a target width
\newcolumntype{Y}{>{\raggedright\arraybackslash}X}

\usepackage{tikz}
\usepackage{pgfplots}
\pgfplotsset{compat=1.18}
\usepgfplotslibrary{groupplots}
\newcommand{\edit}[1]{#1}
\newcommand{\editU}[1]{#1}
\newcommand{\editG}[1]{#1}

\usepackage{xspace}

\newcommand{\comm}[1]{}

\newcommand{\declarativebaseline}{DeclBase}

\begin{document}
%\title{Procedural Scene Programs for Open-Universe Scene Generation: LLM-Free Error Correction via Program Search}
\title[Procedural Scene Programs]{Procedural Scene Programs for Open-Universe Scene Generation: LLM-Free Error Correction via Program Search}

\author{Maxim Gumin}
\email{maxim_gumin@brown.edu}
\affiliation{%
    \institution{Brown University}
    \country{USA}
}

\author{Do Heon Han}
\email{do_heon_han@brown.edu}
\affiliation{%
    \institution{Brown University}
    \country{USA}
}

\author{Seung Jean Yoo}
\email{seung_jean_yoo@brown.edu}
\affiliation{%
    \institution{Brown University}
    \country{USA}
}

\author{Aditya Ganeshan}
\email{aditya_ganeshan@brown.edu}
\affiliation{%
    \institution{Brown University}
    \country{USA}
}

\author{R. Kenny Jones}
\email{russell_jones@brown.edu}
\affiliation{%
    \institution{Brown University}
    \country{USA}
}

\author{Kailiang Fu}
\email{kailiang_fu@brown.edu}
\affiliation{%
    \institution{Brown University}
    \country{USA}
}

\author{Rio Aguina-Kang}
\email{raguinakang@ucsd.edu}
\affiliation{%
    \institution{UC San Diego}
    \country{USA}
}

\author{Stewart Morris}
\email{stewart_morris@brown.edu}
\affiliation{%
    \institution{Brown University}
    \country{USA}
}

\author{Daniel Ritchie}
\email{daniel\_ritchie@brown.edu}
\affiliation{%
    \institution{Brown University}
    \country{USA}
}

\begin{abstract}
Synthesizing 3D scenes from open-vocabulary text descriptions is a challenging, important, and recently-popular application.
One of its critical subproblems is \emph{layout generation}: given a set of objects, lay them out to produce a scene matching the input description.
Nearly all recent work adopts a \emph{declarative} paradigm for this problem: using an LLM to generate a specification of constraints between objects, then solving those constraints to produce the final layout.
In contrast, we explore an alternative \emph{imperative} paradigm, in which an LLM iteratively places objects, with each object's position and orientation computed as a function of previously-placed objects.
The imperative approach allows for a simpler scene specification language while also handling a wider variety and larger complexity of scenes.
We further improve the robustness of our imperative scheme by developing an error correction mechanism that iteratively improves the scene's validity while staying as close as possible to the original layout generated by the LLM.
In forced-choice perceptual studies, participants preferred layouts generated by our imperative approach 82\% and 94\% of the time when compared against two declarative layout generation methods.
We also present a simple, automated evaluation metric for 3D scene layout generation that aligns well with human preferences.

\end{abstract}

\begin{CCSXML}
<ccs2012>
   <concept>
       <concept_id>10010147.10010371</concept_id>
       <concept_desc>Computing methodologies~Computer graphics</concept_desc>
       <concept_significance>500</concept_significance>
       </concept>
   <concept>
       <concept_id>10010147.10010257.10010293.10010294</concept_id>
       <concept_desc>Computing methodologies~Neural networks</concept_desc>
       <concept_significance>500</concept_significance>
       </concept>
   <concept>
       <concept_id>10010147.10010178.10010179.10010182</concept_id>
       <concept_desc>Computing methodologies~Natural language generation</concept_desc>
       <concept_significance>500</concept_significance>
       </concept>
 </ccs2012>
\end{CCSXML}

\ccsdesc[500]{Computing methodologies~Computer graphics}
\ccsdesc[500]{Computing methodologies~Neural networks}
\ccsdesc[500]{Computing methodologies~Natural language generation}

\keywords{scene synthesis, program synthesis, layout generation, large language models}

\begin{teaserfigure}
    \centering
    \setlength{\tabcolsep}{2pt}
    \newcolumntype{y}{>{\centering\arraybackslash}p{0.24\linewidth}}
    \begin{tabular}{yyyy}
         \textit{``A casino''} &
         \textit{``Forest clearing''} &
         \textit{``Depths of hell''} &
         \textit{``A post-apocalyptic campsite''}
         \\
         \includegraphics[width=\linewidth]{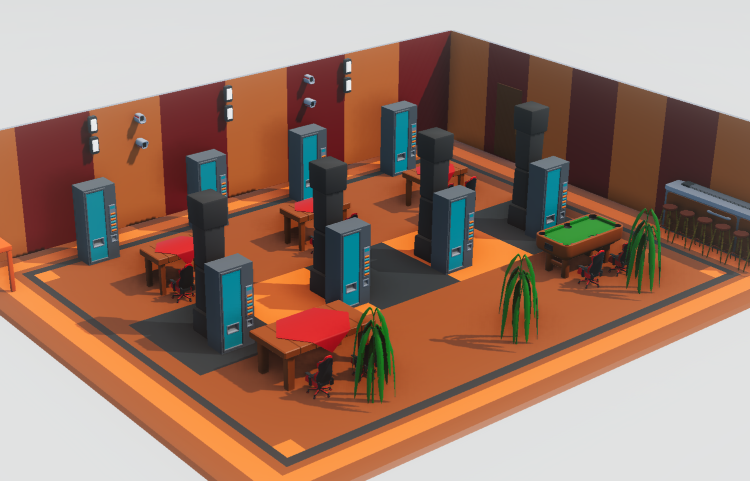} &
         \includegraphics[width=\linewidth]{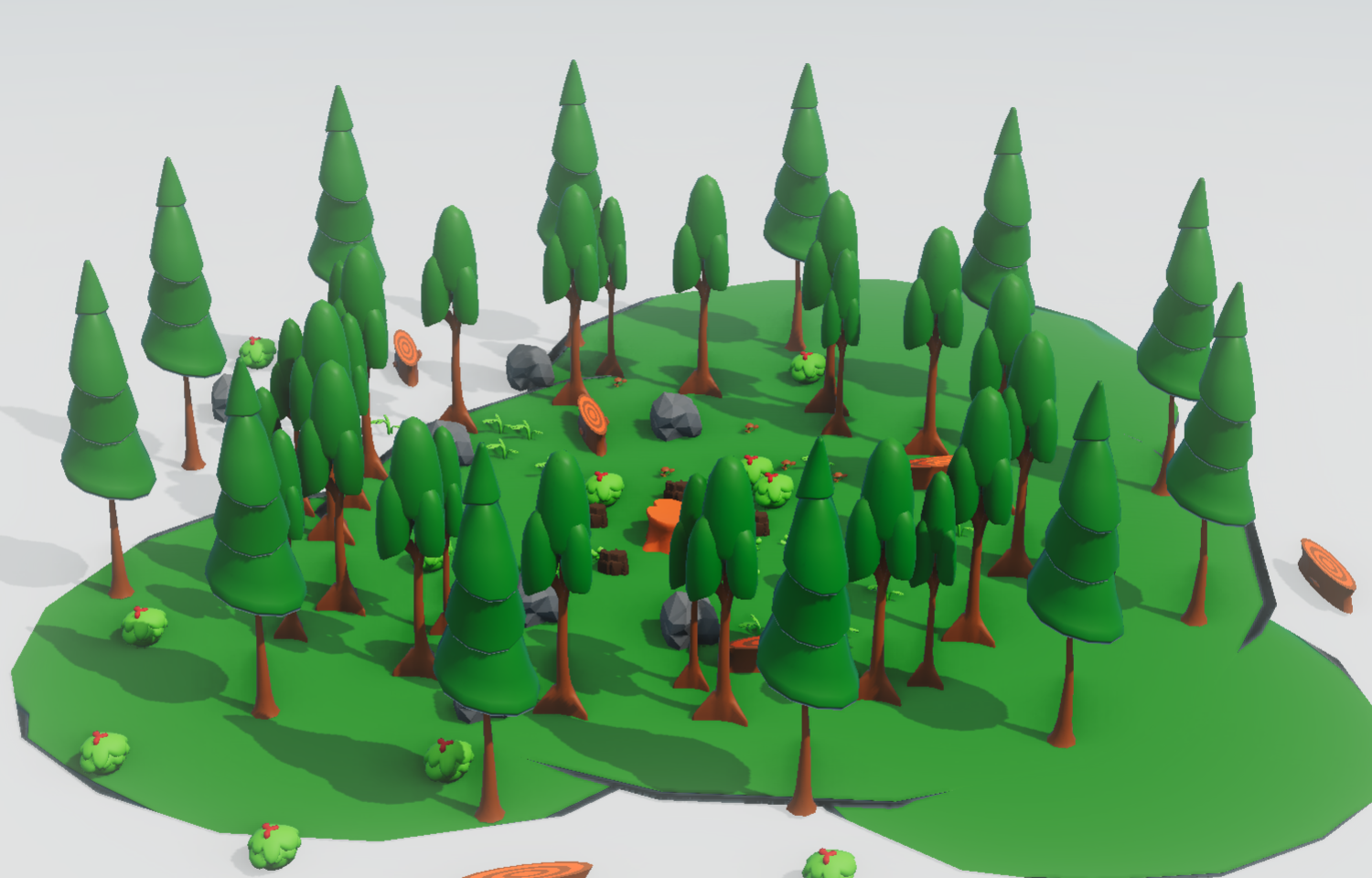} &
         \includegraphics[width=\linewidth]{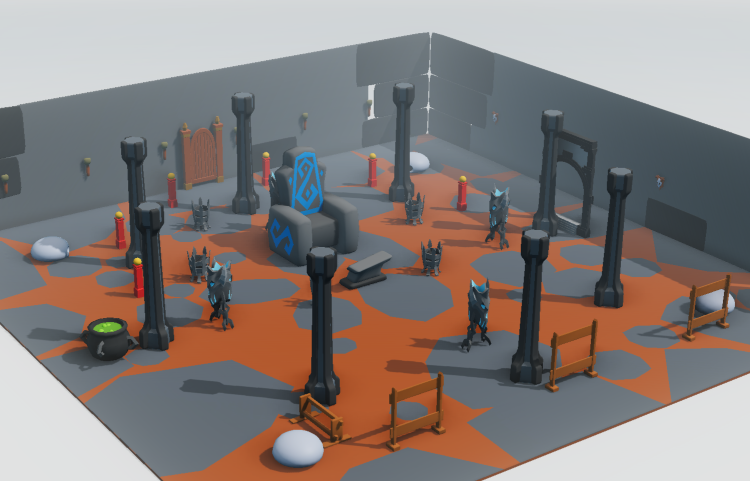} &
         \includegraphics[width=\linewidth]{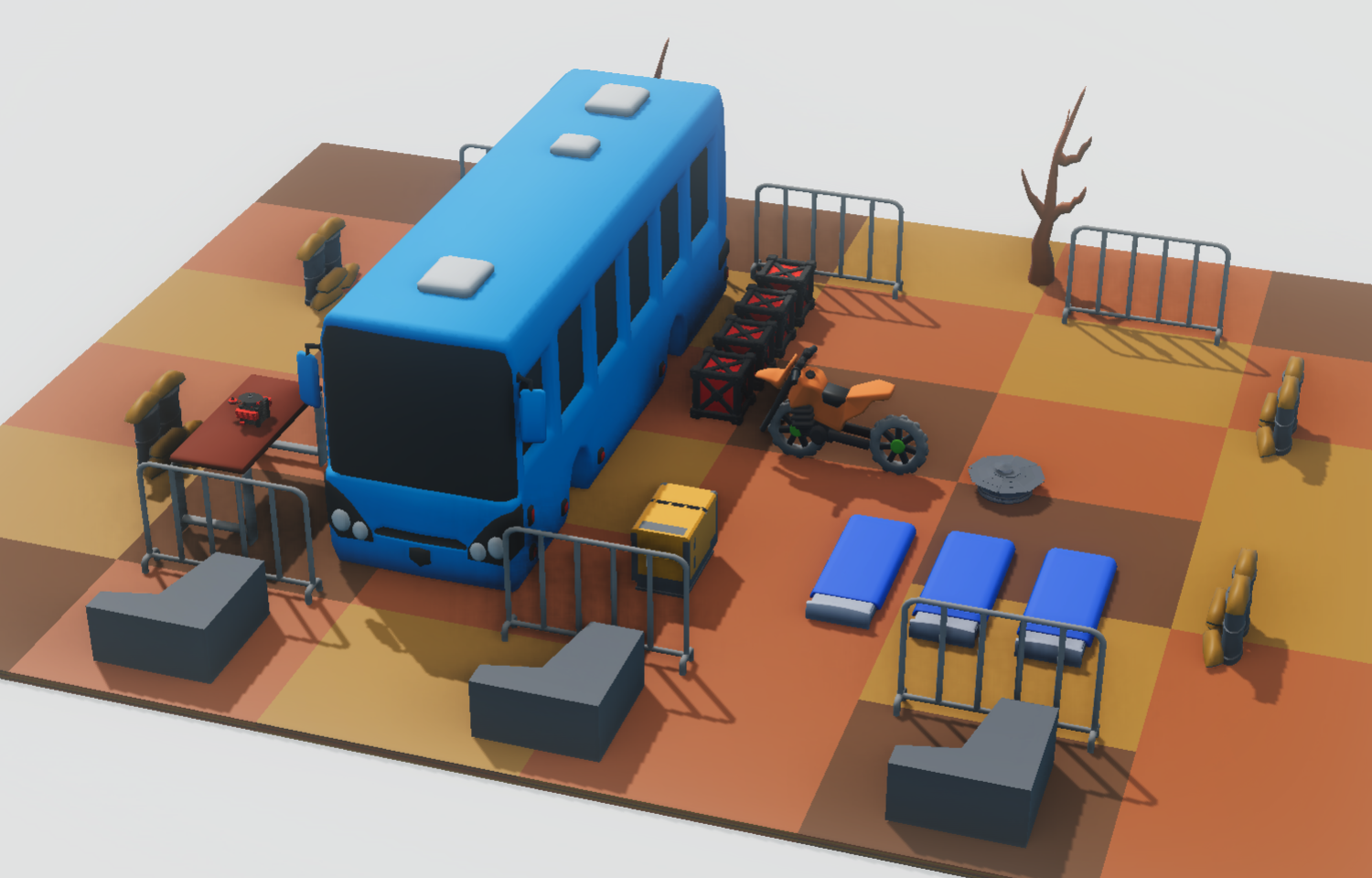}
         \\
         \textit{``Stonehenge''} &
         \textit{``An ice city''} &
         \textit{``A classroom''} &
         \textit{``A graveyard''}
         \\
         \includegraphics[width=\linewidth]{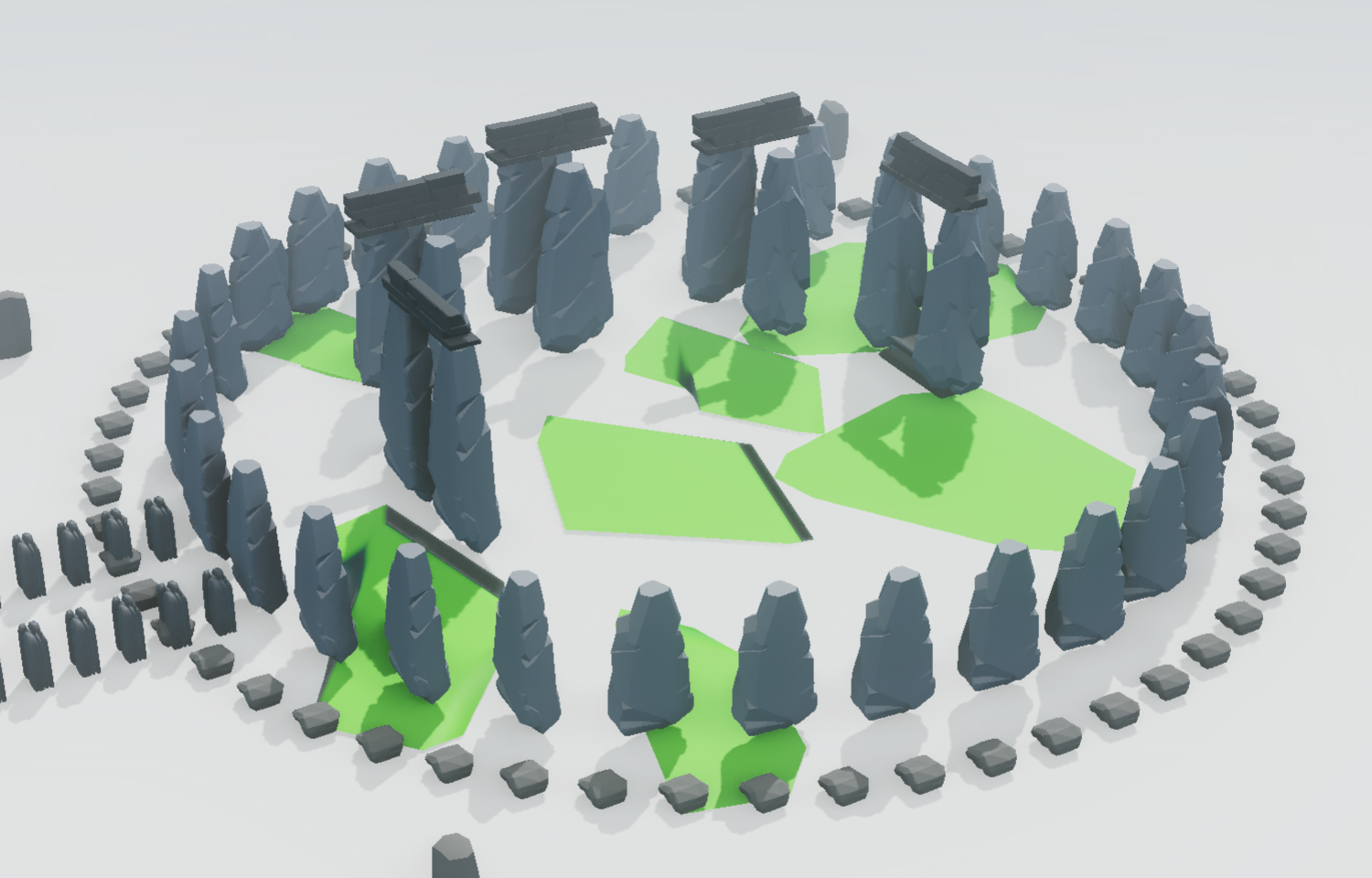} &
         \includegraphics[width=\linewidth]{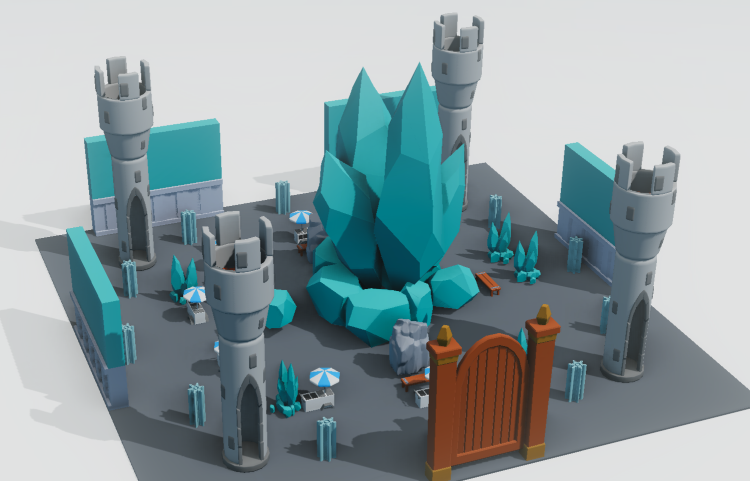} &
         \includegraphics[width=\linewidth]{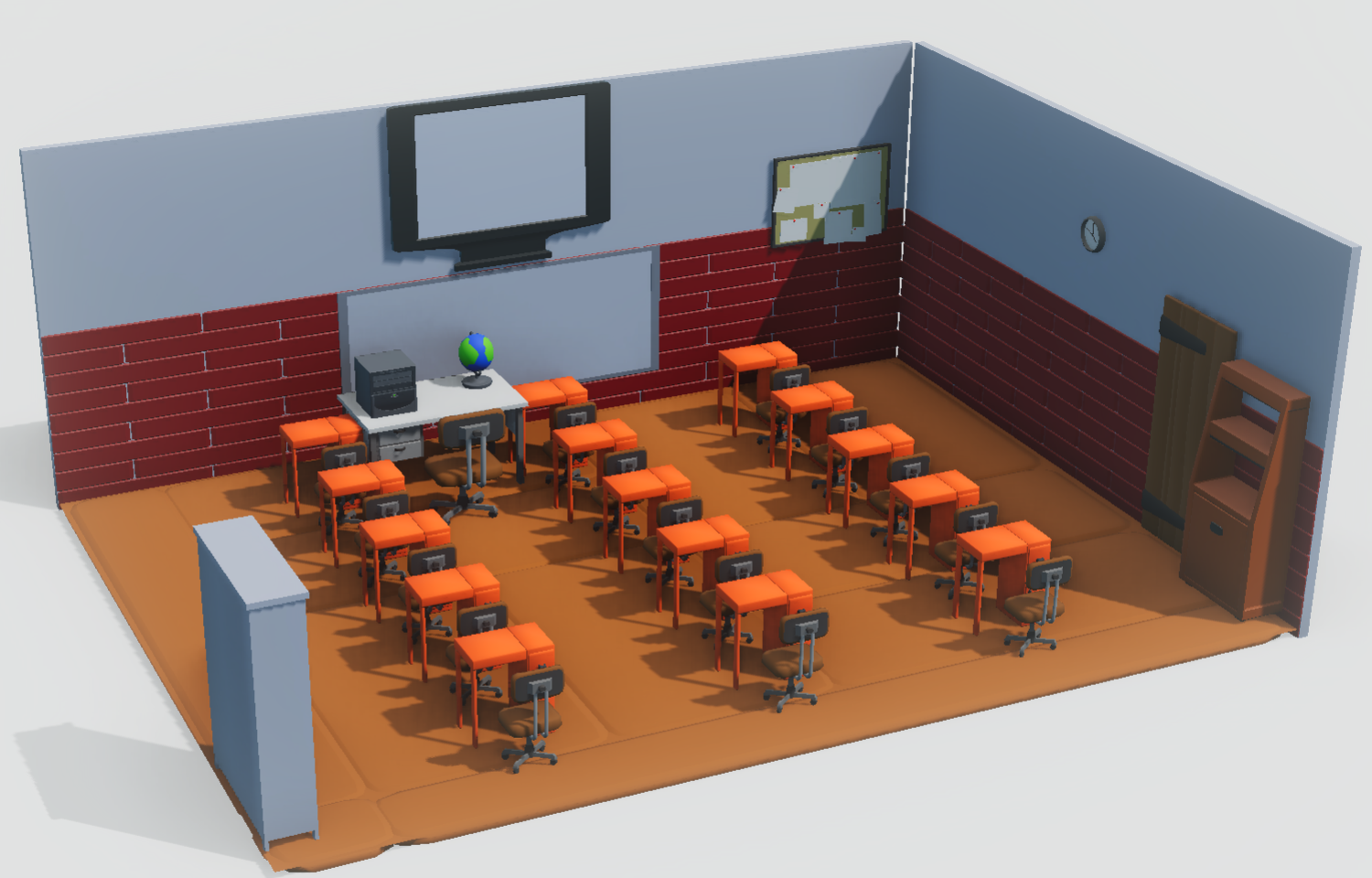} &
         \includegraphics[width=\linewidth]{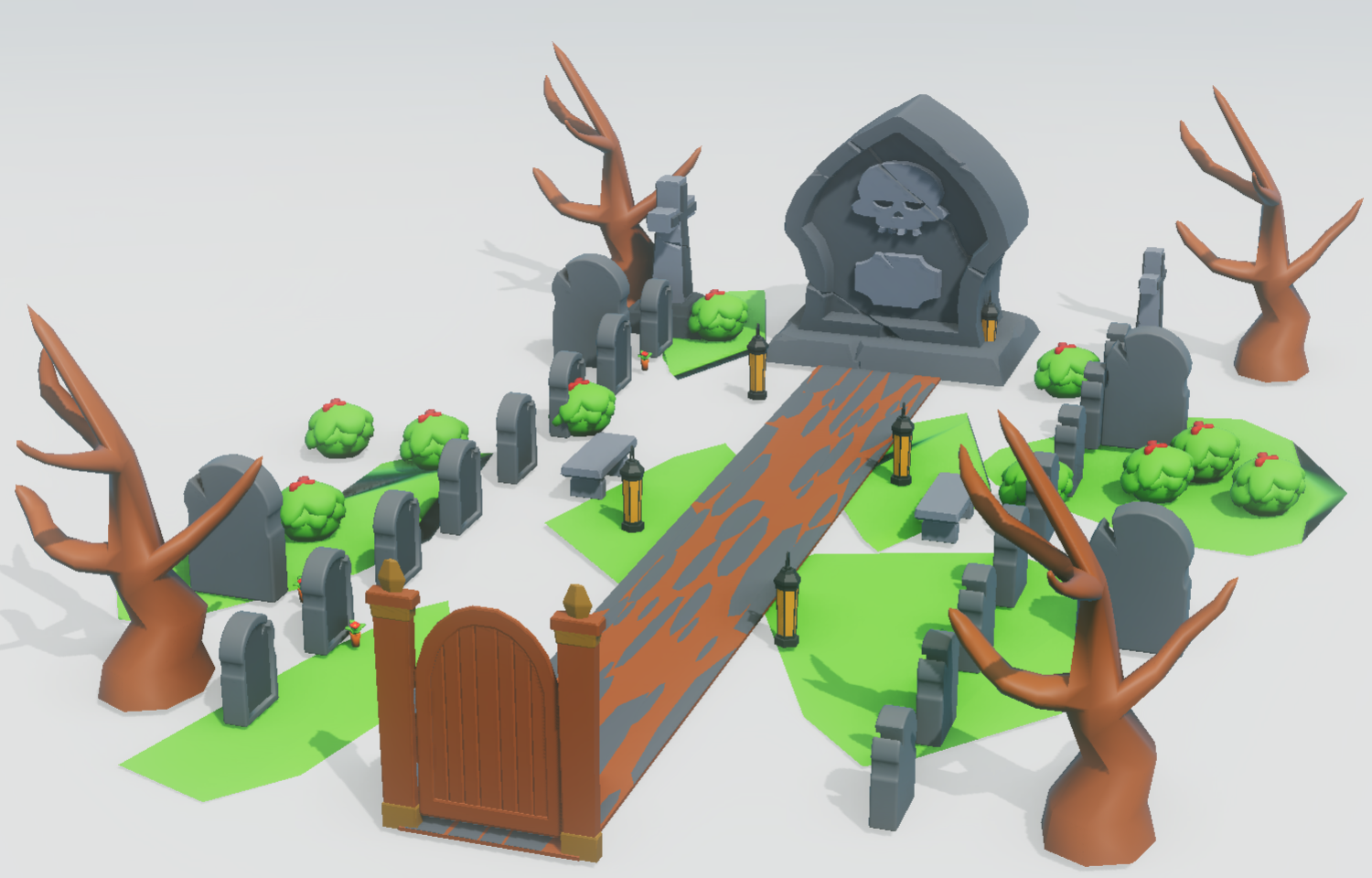}
         %\textit{``Railway station platform''} &
         %\textit{``Chessboard''} &
         %\textit{``Kitchen for a Family of 12''} &
         %\textit{``Wizard's laboratory''}
         %\\
         %\includegraphics[width=\linewidth]{figs/new_teaser/railway.png} &
         %\includegraphics[width=\linewidth]{figs/new_teaser/chess.png} &
         %\includegraphics[width=\linewidth]{figs/new_teaser/kitchen.png} &
         %\includegraphics[width=\linewidth]{figs/new_teaser/wizards.png}
    \end{tabular}
    \caption{
        Our method generates 3D indoor and outdoor scene layouts from open-ended text prompts. Generated layouts are not limited to a fixed set of room types or object categories.
        All layouts in the figure are generated as procedural programs, where the LLM's mistakes are corrected using our program search mechanism.
    }
    \label{fig:teaser}
\end{teaserfigure}

\maketitle
\renewcommand{\shortauthors}{Gumin et al.}
%\begin{center}Paper ID: papers\_1820s1\end{center}

% \kenny{QUESTIONS:}
%
% Lightweight models? Not sure how this fits narratively
% Simpler LLMs?

% error-correction: coordinate descent?
% taxonomy of scenes?
% what else for results? 

\section{Introduction}

%The area I’m working on and why it’s important.
%The specific problem(s) I’m tackling.

3D scenes serve as representations of the environments surrounding us: homes, workplaces, social gathering spaces, etc. They can also represent virtual worlds for games, films, and architecture.
In this paper, we study the problem of \emph{open-universe scene generation}: synthesizing a 3D scene from a natural language prompt, where prompts are not limited to a fixed vocabulary, and objects are not restricted to a fixed set of object categories. 

Large language models (LLMs) are a natural fit for this task given their vast knowledge bases.
A 3D scene can be viewed as a collection of objects, where each object is specified by attributes such as size, mesh, position, and orientation.
Synthesizing such a scene involves several steps: generating a set of objects, determining the positions and orientations of those objects (i.e. layout), and generating or retrieving 3D meshes for each object.
In this paper, we focus on the layout subproblem.
    
%Why those problems remain tricky with current technologies (i.e. why existing state-of-the-art isn’t quite enough).

The first works that studied LLM-based scene generation took an \textit{imperative} approach, directly specifying the desired layout state.
\edit{LayoutGPT}~\cite{feng2023layoutgpt} is a representative example, where an LLM designs a layout by  
iteratively predicting object parameters (sizes, positions, orientation).
This paradigm offered a number of advantages, as LLMs showed an impressive ability to know what objects should populate a scene across a wide range of scenarios.
While this allowed LLMs to quickly produce 3D scene layouts from text, 
LLMs struggled with accurate placement of objects and would produce layouts that violated physical properties (overlapping objects, floating objects)~\cite{makatura2023large}.

% \aditya{Maybe we don't need to have the timeline laid out anymore (earlier was imperative then decl and now back to imperative). Perhaps we can just have - there are declarative approaches and they have these problems. The alternative is imperative - it solves some of these issues but has other. We show that these others can be solved by using EC. By solving these issues with imperative we end up designing a superior system.}

To overcome these limitations, \edit{recent work has overwhelmingly adopted \emph{declarative} approaches}~\cite{yang2023holodeck,SceneCraft,wen2023anyhome,aguinakang2024openuniverseindoorscenegeneration,çelen2024idesign,kodnongbua2024zeroshotsequentialneurosymbolicreasoning,Littlefair2025_FlairGPT}. 
Unlike the imperative paradigm, these declarative specifications do not directly specify object attributes, but rather describe properties that the layout should have.
These operations are often designed to associate with object-to-object relations such as \texttt{on(a,b)}, \texttt{adjacent(a,b)} or \texttt{aligned(a,b,c)}. 
An execution module is then used to convert these programs into a layout, e.g. a solver can search for a layout that best minimizes the error with respect to the LLM-predicted relations.
A nice property of this paradigm is that instead of predicting object parameterizations directly and independently, the LLM's job is simplified, as its prediction can use higher-level operations.
%The rationale behind the declarative paradigm is that it should be easier for an LLM to reason about sentences such as “the lamp is on the table” or “the chair is adjacent to the table” than about precise numeric values.
%\kenny{Natural language analogs?}

While declarative approaches show \edit{advantages over simple imperative schemes like LayoutGPT}~\cite{yang2023holodeck,aguinakang2024openuniverseindoorscenegeneration}, this framing is not without its own limitations. Declarative approaches can be slow to execute, as they have to solve a complex optimization problem just to produce a single layout. This is especially true for scenes containing a large number of objects (e.g. museums, theaters), because the time required for a solver to find a satisfying configuration of object positions depends at least linearly on the
number of relations, and the number of relations usually depends quadratically on the number of objects.
Such slowness makes it more difficult to edit or modify a scene.
In addition, not every object configuration can be expressed in a particular declarative domain-specific language (DSL). For example, it can be hard to arrange objects in a spiral if the declarative DSL has no explicit \texttt{spiral()} constraint.
%Declarative approaches find layouts that minimize some error with respect to chosen constraints, but this scales poorly as scenes become more complex.
%Specifying up-front every single constraint you care about can work for simple scenes with few objects, but without the right language operations, this can become unwieldy for larger scenes (e.g. grocery stores, city blocks). 

In searching for ways to improve LLM-based scene generation, we can take inspiration from how declarative approaches made improvements upon early imperative methods.
So, what lead to the dominance of the declarative paradigm?
For one, declarative programming languages used higher-level operations that had more natural semantic analogs, which simplified the LLM prediction task.
Perhaps more importantly, declarative paradigms have a built-in error-correction module---their solver.
That is, all existing LLM models make mistakes when producing scene layout specifications, but solving for an error-minimizing layout can mitigate this issue.

Are the above differentiators unique to the declarative setting, or could they be realized in an imperative approach as well?
The imperative programming paradigm has a number of benefits over the declarative formulation.
Execution is fast, direct and exact (specifying a single scene). 
Declarative programs can only enforce relationships encoded as constraints in their domain-specific language, while imperative programs in principle have the capacity to realize any possible object configuration. 
How can we take the lessons from the declarative paradigm and use them to improve the imperative approach?

We propose a co-designed system for LLM scene synthesis that combines procedural scene programs with a symbolic error-correction module. 
To this end, we introduce a new relation-centric design pattern for procedural scene programs, which we implement in a Python-embedded DSL we call Procedural Scene Description Language (PSDL).
Beyond simply specifying each object position as an independent decision, PSDL allows object attributes to be defined using \textit{parametric relationships} with respect to the attributes of previously defined objects or constants.
PSDL also includes a domain-specific operator for specifying \textit{local coordinate frames}, simplifying the task for the LLM, as it can create sub-layouts in a canonical space that are then cohesively transformed into the global scene. 
Since PSDL programs are embedded in Python, they allow the LLM to make use of \textit{control flow} operations such as loops and conditionals, supporting the definition of complex layout patterns.

In tandem with the design of PSDL, we co-design an error-correction scheme that makes use of these improved procedural programs.
Our error correction module iteratively refines LLM-generated programs while preserving their original structure as much as possible. 
The goal of this module is to search for programs that minimize errors in the scene produced by executing the program while also minimizing deviation from the initial program definition.
As our procedural scene representation avoids global solvers, program evaluation is cheap, so our search procedure is able to execute modified programs, observe a new layout, and use this information to decide how to adjust the symbolic scene representation.
We find that a simple local search method performs well, and is particularly effective for the procedural scenes that use our new design patterns.
For instance, due to the presence of shared variables, the error correction module can easily coordinate updates across multiple objects, such as adjusting the layout of an entire row of objects with a single update.

%Our error-correction is fast and cheap to run, only reasoning over the LLM produced scene program, and not requiring extra reasoning steps or additional LLM calls.

%How I evaluate the new approach (i.e. how to convince the reader that this new approach is valuable).

We evaluate our approach against prior imperative and declarative scene generation schemes.
We propose a taxonomy of scene prompts along a number of different axes: small indoor vs large outdoor, common spaces to fantastical scenes, chaotic to highly structured. 
While we find that our new formulation is competitive in all of these settings, it especially excels for prompts describing large, highly-structured scenes.
As these perceptual studies don't scale well, we also introduce a new vision-language model (VLM)-based evaluation scheme that we find is better aligned with human judgments than alternative automatic evaluation approaches that use VLMs.
%Under this new metric we also find that our approach offers improvements, and we ablate decisions in our co-design of our procedural programming paradigm for error-correction. 

In summary, our contributions are:
\begin{enumerate}
    %\denselist
    % \item Developing the imperative paradigm for open-universe scene generation and comparing it against the best systems that follow the declarative paradigm.
    \item The PSDL language for imperative specification of open-universe scene layouts.
    \item An error correction method for PSDL programs that does not involve additional calls to an LLM.
    \item A protocol for evaluating open-universe scene layout synthesis systems, including a benchmark set of input descriptions covering a wide variety of possible scenes.
    \item A human-aligned automated evaluation method for scene layout generation.
\end{enumerate}

\section{Related Work}
\paragraph{Scene Synthesis pre-LLMs}
The problem of scene synthesis has a rich history in computer graphics.
Early work focused on laying out objects based on manually-defined design principles~\cite{ifurniture_design}, simple statistical relationships between objects extracted from a small set of examples~\cite{yu2011MakeItHome}, or with programmatically-specified constraints~\cite{rj_mcmc}.
Later research focused on data-driven methods~\cite{fisher2012,kermani2016learning,liang2017,qi2018human}, with a surge in activity as deep neural networks gained popularity~\cite{FastSynthCVPR,DeepSynthSIGGRAPH2018,wang2019planit,GRAINS,zhou2019scenegraphnet,zhang2019hybrid,Paschalidou2021NEURIPS,wang2020sceneformer,tang2023diffuscene,lin2024instructscene}.
These prior works develop closed-universe generative models (i.e. restricted to certain scene and object categories), and all of them require (in some cases quite large) datasets of 3D scenes for training.
By contrast, LLMs offer the capability---in theory---to synthesize arbitrary types of scenes and to do so with no explicit 3D scene training data.

\paragraph{Scene Synthesis with LLMs}
While research on text-based scene generation predates the rise of LLMs~\cite{WordsEye,chang2017sceneseer3dscenedesign}, their development has led to a new generation of text-to-scene generative models which are both more flexible and more open-ended than earlier systems.
LayoutGPT~\cite{feng2023layoutgpt} was a pioneering system that operated in a very simple imperative paradigm by iteratively specifying object coordinates, dimensions, and orientations. 
This `coordinate-centric' framing has seen adoption in other scene generation framings as well:~\cite{bhat2025cube,ocal2024sceneteller}.
The Scene Language~\cite{zhang2024scenelanguagerepresentingscenes} is a concurrent work that also uses an imperative approach for producing scene layouts of objects, which can then be converted into detailed scene through a neural rendering scheme.
This system produce scene programs that use control flow, but doesn't use parametric relationships, local coordinate frames, or any error-correction mechanism.
Many other LLM scene generation works have adopted a declarative approach, i.e. using an LLM to produce a declarative program specifying the layout constraints~\cite{yang2023holodeck,SceneCraft,wen2023anyhome,aguinakang2024openuniverseindoorscenegeneration,çelen2024idesign,kodnongbua2024zeroshotsequentialneurosymbolicreasoning,Littlefair2025_FlairGPT}.
We experimentally compare the benefits of our procedural formulation with an integrated error-correction module over these alternatives.

%Our work differs from these approaches, as we employ an imperative approach to scene synthesis, i.e. the LLM generates explicit instructions for constructing scenes by iteratively placing objects with relative positioning.
%In addition, we present an apples-to-apples comparison between declarative and imperative approaches to scene layout generation, which we expect to be instructive for designing DSLs for LLM-based program synthesis on other tasks such as 3D shape modeling and editing.

\paragraph{Automatically Correcting LLM Outputs}
% What are some methods which don't go on self correction?
As LLMs can fail to produce correct output in one shot, many prior works deploy corrective mechanisms to refine LLM-generated outputs, including some work on LLM-based scene generation~\cite{SceneCraft}. 
The most common, and general, approach is self-correction, where the output of an LLM is iteratively refined by the LLM itself~\cite{pan2023automaticallycorrectinglargelanguage}.
Such self-correction mechanisms, while appealing in theory, are costly to run (requiring multiple LLM calls), and on code generation tasks, they typically offer modest or no real performance gain~\cite{olausson2024repair}.
Instead, we propose an efficient error correction scheme based on iterative local search, finding programs that are close to the LLM's original output while minimizing errors such as object overlaps.
We compare our proposed error-correction module against LLM self repair, a gradient descent baseline, and alternative formulations, demonstrating its benefits experimentally. 

\section{Overview}

\begin{figure*}[t!]
    \centering
    % \includesvg[width=0.9\linewidth]{Code_Overview}
    \includegraphics[width=0.9\linewidth]{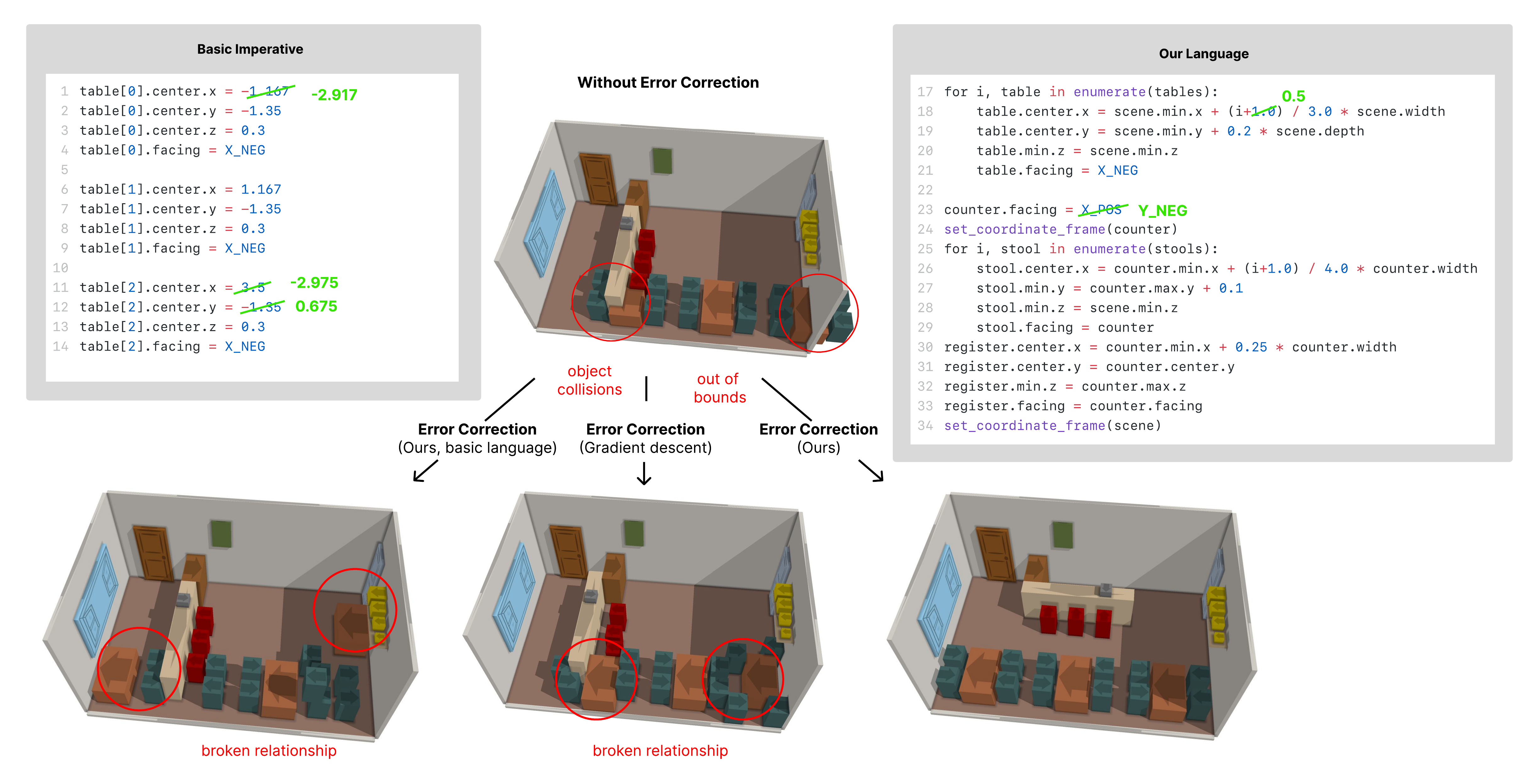}
    \caption{Here we show how our error correction mechanism works for the same initial object configuration expressed in 2 languages: Layout-GPT style basic imperative language (left), and PSDL (right). The initial configuration has a positive loss because some objects go out-of-bounds, and some pairs of objects are overlapping. All corrected scenes have loss zero. However, the corrected version expressed in the basic imperative language, and the scene corrected with gradient descent (middle) break the relationship between tables and chairs. Our experiment shows that, given 2 corrections of the same initial object configuration, the automatic evaluator prefers the corrected version expressed in PSDL in 71.4\% of cases against LayoutGPT-style, see Table~\ref{tab:ec_ablation}.}
    \label{fig:overview}
\end{figure*}

Our goal is to investigate the quality of layouts generated by our method, in comparison to prior approaches. As much as possible, we would like this comparison to show the relative merits of each system for layout generation, rather than details of how a particular scene synthesis system was implemented. Thus, we design a scene synthesis pipeline which factors out computational stages not relevant to layout generation and shares those stages in common between the different layout generation methods that we consider.

In the first stage, an LLM takes a textual scene description as input and outputs a ``scene template'' consisting of the dimensions of the scene and a list of objects. Each object is defined by a name, a set of dimensions, and one of three types of physical support: \texttt{STANDING}, \texttt{WALL-MOUNTED}, or \texttt{FLOATING}. \editG{This template is then passed to the layout generation stage, which determines each object’s position and orientation.}

\editG{
In the first sub-stage of layout generation, an LLM writes a scene program in our Python-embedded DSL (PSDL). The LLM is prompted with a system prompt that says, in effect: given a scene name and the list of objects, produce a high-quality layout described in PSDL. The prompt explains how to use the DSL and emphasizes the core principle: place new objects relative to already placed objects or to the scene cuboid. It also encourages variable sharing to minimize repeated numeric constants (e.g., prefer a single parameter reused across placements rather than duplicating the same literal). Four in-context examples (provided in the supplement) accompany this system prompt to illustrate the intended use of the DSL.

In the second sub-stage of layout generation, an LLM-free error-correction module rewrites parts of the generated scene program to remove layout violations while preserving the program’s structure. The layout generation stage differs across the systems we evaluate, but all share the same upstream template and downstream visualization components.
}

%This scene template is then passed to a layout generation stage, which determines the positions and orientations of its objects. Our layout generation consists of two sub-stages: first, an LLM writes a scene program (prompted with a system prompt and four in-context examples, provided in the supplementary material); second, an LLM-free error correction stage rewrites parts of the scene program. The layout generation stage differs across the systems we evaluate.

% (ours with different error-correction variants, Holodeck, FlairGPT)

%This scene template is then passed to a layout generation stage to determine the positions and orientations of its objects. Layout generation stage is different for each system being evaluated (ours with different variants of error correction, Holodeck, FlairGPT).

Finally (and optionally), the pipeline can invoke an object retrieval stage to retrieve a 3D mesh for each object in the layout. Object retrieval is not the focus of our work, but we include it in our pipeline for visualizing some qualitative results. In our implementation, we use simple CLIP similarity~\cite{clip} to retrieve a mesh whose rendered image matches the object's name. We retrieve meshes from the HypeHype Asset Library~\cite{HypeHypeAssets}, which currently contains about 6000 3D model assets. Other 3D shape datasets could also be used here~\cite{chang2015shapenet,objaverse,objaverseXL}.

By comparing different layout methods while keeping the template synthesis and object retrieval stages the same, we ensure that scenes being compared have the same size and the same set of objects, making the comparison fair. To be consistent with the prior methods to which we compare, we restrict object orientations to four cardinal directions.

\section{Procedural Scene Description Language}
\label{sec:psdl}

In this section, we present our new scene description language and motivate why its various features make it more amenable to error correction.

On one hand, existing declarative scene layout methods like Holodeck~\cite{yang2023holodeck} rely on expensive global solvers, making search-based error correction loop infeasible time wise. On the other hand, existing imperative methods like LayoutGPT~\cite{feng2023layoutgpt} express scenes as a flat list of object placements instructions. As a result, the search to repair such programs can only be conducted over the space of every object’s position and orientation without regard for the higher-level relations that are critical for the scene. Consequently, symbolic repair of such programs often eliminates spatial violations at the cost of breaking critical scene relations.

As a solution to this predicament, we propose a new language, Procedural Scene Description Language (PSDL), which has two benefits. First, it is an imperative language which helps to avoid expensive constraint solvers. Second, PSDL provides language constructs which help express a variety of scene relations procedurally, which makes performing error correction on PSDL programs stronger. Specifically, PSDL distinguishes itself from prior imperative languages through three key features:

\vspace{-0.2em}
\paragraph{Explicit Object Relationships} Object positions and orientations in PSDL are not simple numerical values. Instead, they are functions defined relative to previously placed objects or scene bounds (see Table~\ref{tab:imperative-dsl-features} left). Because these expressions are preserved during repair, object–to-object relations survive most parameter perturbations (see Fig. \ref{fig:overview}).

\vspace{-0.2em}
\paragraph{Expression sharing through variables} Variables allow a single symbolic constant to control many objects (see Table~\ref{tab:imperative-dsl-features} right). Changing one parameter --- say, the spacing between chairs --- coherently updates all dependents, drastically shrinking the space the search must explore.

\vspace{-0.2em}
\paragraph{Control-flow for structured repetition} Loops and conditionals let programs capture patterns (rows of desks, symmetric place settings, etc.) concisely. A loop introduces implicit sharing: the constant stride inside the loop is automatically reused across all iterations, again tightening the search domain (see Table~\ref{tab:imperative-dsl-features} right).

%\vspace{0.5em}
Together these features enable compact expression of scenes while benefiting from the solver-free execution of imperative programs. Representing the scene procedurally allows us to conduct the repair search at the level of parametrizations of the procedural program, rather than at the object level. Consequently, every candidate program variant preserves the scene relations by construction. Hence, our search space is compact and more semantically meaningful, which makes fixing violations less likely to disrupt essential scene structure. We evaluate the effect of different PSDL features on the quality of generated scenes in Section~\ref{sec:results}.

\subsection{Language Details}
\editU{Semantically, a PSDL program takes a scene template (scene dimensions and a pre-instantiated list of objects) as input and returns positions and orientations for those objects. Implementation-wise, the template supplies a short prelude that is concatenated with the PSDL code, binding object identifiers and setting the scene dimensions; the program then executes in this environment. The result is conveyed by side effects, namely updates to object attributes.

To make PSDL practical and expressive, we implement it as an embedded domain-specific language in Python. PSDL programs are allowed to be fully arbitrary Python code that additionally leverages the provided domain-specific operators; accordingly, generated programs may freely use standard library functions such as \texttt{cos}, \texttt{enumerate}, \texttt{filter}, \texttt{isinstance}, \texttt{len}, \texttt{map}, \texttt{random}, \texttt{range}, \texttt{sum} and \texttt{zip}. In the following paragraphs, we highlight the key domain-specific operators and explain their meaning.}

%Generated programs may use functions from the Python standard library such as \texttt{cos}, \texttt{enumerate}, \texttt{filter}, \texttt{isinstance}, \texttt{len}, \texttt{map}, \texttt{random}, \texttt{range}, \texttt{sum}, \texttt{zip}, etc. in addition to PSDL's domain-specific functions.

\edit{
All scene objects \texttt{o} are represented as cuboids, with dimensions \texttt{o.width}, \texttt{o.depth}, and \texttt{o.height}. Width is defined as the dimension perpendicular to the object's facing direction, while depth is defined along the facing direction. Each object exposes its geometric center \texttt{o.center}, its orientation relative to the current coordinate frame \texttt{o.facing},} \editG{which takes values among} \edit{\texttt{X\_NEG}, \texttt{X\_POS}, \texttt{Y\_NEG} or \texttt{Y\_POS}, and its axis-aligned bounding box (AABB). Object's AABB is defined by two 3d vectors: \texttt{o.min} contains the minimum \texttt{x}, \texttt{y}, and \texttt{z} coordinates (in the current coordinate frame) of the box, while \texttt{o.max} contains the maximum. The special \texttt{scene} object represents the bounding cuboid of the entire scene.

By default, the program's coordinate frame is centered on the scene cuboid, but it can be re-centered on a particular object \texttt{o} using \texttt{set\_coordinate\_frame(o)}, which aligns the y-axis with the object’s facing direction, the x-axis $90^\circ$ clockwise from y, and the z-axis upward. 
For expressions like \texttt{chair.max.x = table.min.x - 0.1} (see Table}~\ref{tab:imperative-dsl-features}\edit{) to change not only the \texttt{chair.max} vector, but the position of the \texttt{chair} itself, we overload Python's assignment operator.}

An exhaustive PSDL API is provided in Supplementary Tables~\ref{tab:psdl-types},~\ref{tab:psdl-ops}.

% While we have not proved it, we think that the findings of the paper generalize to languages that follow PSDL’s design patterns (explicit object relationships, expression sharing between objects, control flow), and not just to our specific DSL.

\begin{table}[t!]
\centering
\caption{Key features of PSDL. The left column demonstrates explicit geometric relationships for positioning objects relative to others. The right column shows the use of variables to define reusable patterns, enabling concise scene descriptions. Together, these features allow PSDL to describe scenes precisely and efficiently.}
\label{tab:imperative-dsl-features}
\begin{tabular}{l|l}
\toprule
\small \textbf{Explicit Geometric Relationships} & \small \textbf{Use of Loops and Variables} \\ 
\hline
\small \texttt{chair.max.x = table.min.x - 0.1} & \small \texttt{d = 2.0} \\
\small \texttt{chair.center.y = table.center.y} & \small \texttt{for i, c in enum(cols):} \\
\small \texttt{chair.min.z = scene.min.z} & \small \texttt{\quad c.center.x = \textbackslash } \\
\small \texttt{chair.facing = table} & \small \texttt{  scene.center.x +  i * d} \\
\bottomrule
\end{tabular}
\end{table}

\section{LLM-Free Error Correction}
\label{sec:errorcorrection}
%Now, given an imperative scene where program evaluation is cheap, the question is how to perform error correction? The goal is to remain as close as possible to the original program (or retain as much of the semantics of the scene as laid out by the LLM), while removing errors in the scene (which kind of errors?). To do this, we formulate an error correction mechanism balances proximity to original program which resolving errors in the scene. In  section~\ref{sec:errorcorrection}, we present this.

%The previous section presented our new procedural language and motivated why its various features make it more amenable to error correction. Now, we present our approach to automatically correcting errors in programs LLMs write using this language. Key goal is to avoid additional calls to LLMs.

We now describe our approach for correcting layout errors in PSDL programs. Although LLMs are capable of generating these programs, LLMs often make errors. We classify LLM errors as either \emph{exceptions} or \emph{layout errors}.

Exceptions include hallucinated function calls, incorrect argument usage, or ``index out of range'' errors that trigger exceptions at runtime. When such an error occurs, we discard the current program and request a new layout from the LLM. Layout errors arise when objects overlap, exceed scene boundaries, or violate support constraints. A common cause of layout errors is an LLM ``forgetting'' previously placed objects; for example, a door might be placed first, then a chair to its left, and, 20 lines of code later, a table to the chair’s right, which blocks the door from opening.

In this paper, we focus on layout errors and seek to correct them without additional calls to the LLM (so-called ``self-repair''), which can be time-consuming, as well as monetarily costly if LLM APIs are used. We first present our formalization of the error correction problem, then we discuss the algorithm we use to solve it.

\subsection{Formulation}
To quantify how problematic a layout is, we define a \emph{loss function} $\text{loss}(L)$ that is composed of the following terms:
\begin{enumerate}
\item \textbf{Out-of-Bounds Loss}: For each object, this loss is the maximum linear distance by which its bounding cuboid protrudes outside the scene boundary. If the object is fully within bounds, this term is zero.
\item \textbf{Overlap Loss}: For each pair of objects, this loss is the cube root of the volume of the intersection of objects' bounding cuboids. To ensure usability, doors and windows are assigned expanded collision boxes to account for opening space and prevent obstruction.
\item \textbf{Standing Loss}: For each \texttt{STANDING} object, this term measures the distance from the bottom face of the object’s cuboid to the nearest horizontal surface that can support it.
\item \textbf{Mounted Loss}: For each \texttt{MOUNTED} object, this term is the distance between the object’s mountable face (usually the back or a side) and the nearest vertical surface that can support it.
\end{enumerate}
These components ensure that the loss function captures both structural integrity and functional correctness in the layout. The same classes of errors are usually represented as default constraints in declarative approaches.

Our goal is to modify the program written by the LLM such that errors are removed (as much as possible), or, equivalently, the $\text{loss}$ function is minimized. However, in fixing errors, modifications to the scene program might destroy the semantic integrity of the scene (e.g. one could fix an overlap between a dining table and its chairs by moving the table to a completely different empty region of the room). Thus, we seek to minimize the errors while maintaining semantic consistency/integrity. Evaluating whether (and to what extent) semantic integrity is preserved is challenging; one could attempt to use an LLM to do so, but one of our explicit goals is to avoid further LLM calls. Instead, we take a conservative approach, leveraging a sufficient-but-not-necessary property: the more similar a modified program is to the original program, the more likely it is to be semantically consistent with that original program. Our goal then becomes: minimize errors while maximizing similarity between the original program and the modified one. We formalize this goal as an optimization problem:
$$\text{argmin}_P [ \text{loss}(L)+d(P,P_0) ]$$
where $P_0$ is the original program, $P$ is the modified program, $L$ and $L_0$ are layouts of $P$ and $P_0$ respectively, and $$d(P,P_0)=d_{\text{edit}}(P,P_0)+d_{\text{OT}}(L,L_0),$$ where $d_{\text{edit}}(P,P_0)$ is the edit distance between programs $P$ and $P_0$, and $$d_{\text{OT}}(L,L_0)=\min_{f\in F}\sum_{o \in O(L)}\text{Vol}(o)\cdot\|\text{center}(o)-\text{center}(f(o))\|_2$$ is the optimal transport distance \cite{peyre2019computationalOT} between layouts. Here $O(L)$ is the sets of objects in $L$, and $F$ is a set of bijections between $O(L)$ and $O(L_0)$ that preserve object categories. Adding the mass transport term ensures that larger objects are penalized more for movement, encouraging the preservation of the positions of major scene elements while allowing smaller objects to be adjusted.

To define the edit distance $d_{\text{edit}}(P,P_0)$ between two PSDL programs, we need to define the set of elementary edits $E(P)$. Then, edit distance $d_{\text{edit}}(P,P_0)$ is the length of the shortest sequence of elementary edits that transforms $P$ into $P_0$. In our experiments, we noticed that, when tasked with generating a layout, LLMs make most mistakes in (1) using correct numeric parameters and (2) setting correct object facing directions. Therefore, to make error correction as efficient as possible, we define the set of elementary edits $E(P)$ to consist of rewrites of constant expressions and rewrites of direction expressions.

%fix an out-of-bounds error for one plant in a row of potted plants by snapping it inside the nearest wall, thereby breaking its alignment with all the other plants in the row).

%(we describe the particular set of edits that we use in the next subsection)

\subsection{Algorithm}
%Even minimizing $\text{loss}(L)$ is already an NP-hard problem, since a special case of it is a bin packing problem.

Minimizing the above objective is a complex search problem in a vast space of programs. Complete enumeration of this space (popular in other program synthesis work) is infeasible. Therefore, we must instead use approximations. Since one of our objectives is to maximize similarity to original scene, we consider search algorithms which start from the original program. Then we iteratively improve this program using local search in the space of PSDL programs.

Precisely, given an LLM-generated program $P_{\text{LLM}}$, we iteratively refine it by constructing a sequence of programs $P_0=P_{LLM}, P_1, P_2, \ldots$: $$P_{i+1}=\mathop{\mathrm{argmin}}_{P\in N(P_i)}f(L),\quad f(L)=\text{loss}(L)+d_\text{OT}(L,L_0),$$ where $L$ and $L_0$ are layouts of $P$ and $P_0$ respectively. Here $N(P)$ is the \emph{neighborhood} of program $P$ relative to the edits $E(P)$: $$N(P)=\{e(P)\;|\;e\in E_\text{fin}(P)\},$$ where $E_\text{fin}(P)$ is finite subset of edits randomly sampled from $E(P)$. \edit{Specifically, $E_\text{fin}(P)$ consists of $10$ edits per each constant expression and $4$ edits per each direction expression. For constants, we multiply a constant by a random variable $\pm 4^Y$, where the sign is chosen uniformly from $\{-1,1\}$, and $Y$ is sampled uniformly and independently in the interval $[-1,1]$; for direction expressions, we include all 4 cardinal directions in $E_\text{fin}(P)$.} The iteration terminates when no available edit decreases the objective $f$ more than some small threshold.

Unlike the declarative approach, which adjusts each object individually, our error-correction strategy modifies expressions in the scene description program. This is especially advantageous for scenes with many objects, where a single floating-point parameter often governs multiple placements (e.g., chair spacing in a theater). By working in this lower-dimensional parameter space, the method both preserves the overall structure of the scene --- ensuring aesthetic consistency --- and reduces computational overhead compared to per-object adjustments. On average, only a few adjustments --- $7.13$ per scene --- are sufficient to resolve errors. See the supplemental material for videos illustrating the error correction process.

In the next section, we evaluate this simple LLM-free iterated local search against the error correction baselines of gradient descent and LLM self-repair.

\begin{table}[t!]
    \renewcommand{\arraystretch}{0.8}
    \centering
    %\small
    \caption{Preference rates for scenes generated using our procedural approach with the error correction module vs. two declarative approaches and a variant without the error correction module in a forced-choice perceptual study.}
    \label{tab:perceptualstudy}
    \begin{tabular}{rccc}
        \toprule
        \textbf{Scene Type} & \textbf{\declarativebaseline } & \textbf{ Holodeck} & \textbf{ Ours w/o EC}
        \\
        \midrule
        Ours w/ EC vs. & 82.9\% & 94.3\% & 74.3\%\\
        Ours w/o EC vs. & 61.8\% &  & -\\
        % \\
        % \midrule
        % Small & 71.43\% & 100\%
        % \\
        % Medium & 82.86\% & 91.43\%
        % \\
        % Large & 90.48\% & 95.24\%
        % \\
        % \midrule
        % Indoor & 81.25\% & 95.83\%
        % \\
        % Outdoor & 86.36\% & 90.91\%
        % \\
        % \midrule
        % Realistic & 84\% & 92\%
        % \\
        % Fantastical & 80\% & 100\%
        % \\
        % \midrule
        % Chaotic & 74.19\% & 96.77\%
        % \\
        % Structured & 89.74\% & 92.31\%
        % \\
        \bottomrule
    \end{tabular}
\end{table}

\begin{table}[t!]
    \centering
    %\small
    \caption{How frequently different automated evaluation metrics agreed with the majority human judgment from our perceptual study.
    Our new method (LLMCompare) is simple and outperforms prior approaches designed for evaluating text-to-image generative models.}
    \label{tab:humanalign}
    \begin{tabular}{cccc}
        \toprule
        \textbf{LLMCompare} & \textbf{LLMCompare (no +/-)} & \textbf{VQAScore} & \textbf{DSG} \\
        \midrule
        77.1\% & 70.0\% & 58.6\% & 50.7\%
        \\
        \bottomrule
    \end{tabular}
\end{table}

\begin{table}[t!]
    \renewcommand{\arraystretch}{0.8}
    \centering
    \small
    \caption{Preference rates using the automated evaluation metric. 'Ours' column includes scenes generated using our 70 prompts. 'Holodeck' column includes 52 scene types from MIT Scenes~\cite{Quattoni2009IndoorScenes} used by Holodeck, and 14 longer prompts from Holodeck's qualitative examples. The 'Complex' column includes all prompts with four or more words. Our method consistently outperforms other approaches across all scene categories and performs well even with complex prompts.}
    \label{tab:scene_comp}
    \begin{tabular}{rccccccc}
        \toprule
        & & & \multicolumn{2}{c}{\textit{Scene Source}} & & \multicolumn{2}{c}{\textit{Prompt Type}}\\
        & \textbf{All} &  & \textbf{Ours} & \textbf{Holodeck} & & \textbf{Simple} & \textbf{Complex} \\
        \midrule
         vs. \declarativebaseline & 76.5\% &  & 77.1\% & 75.8\% &  & 77.8\% & 68.4\% \\
         vs. Holodeck             & 82.4\% &  & 90.0\% & 74.2\% &  & 82.9\% & 78.9\% \\
         vs. FlairGPT             & 88.2\% &  & 85.7\% & 90.9\% &  & 87.2\% & 94.7\% \\
        \bottomrule
    \end{tabular}
\end{table}

\begin{table}[t!]
    \renewcommand{\arraystretch}{0.8}
    \centering
    \footnotesize
    \caption{Preference rates over Holodeck by number of objects using auto metric. '10-20' bin had 9 scenes total, '20-30' bin had 45, '30-40' bin had 37, and '40-beyond' had 42. Our method performs well in all bins, especially in scenes with 40 or more objects.}
    \label{tab:complex_comp}
    \begin{tabular}{rccccccc}
        \toprule
        \textbf{\# Objs per Scene}& \textbf{10--20} & \textbf{20--30} & \textbf{30--40} & \textbf{40->} \\
        \midrule
        % vs. \declarativebaseline & ? & ? & ? & ? & ? & ? & ? \\
        vs. Holodeck & 66.7\% & 80.0\% & 72.9\% & \textbf{95.2}\% \\
        \bottomrule
    \end{tabular}
\end{table}

\begin{table}[t!]
    \renewcommand{\arraystretch}{1.1}
    \centering
    \small
    \caption{Ablation study comparing error correction modules types across preference rate, error count, and run time. The amount of errors and run time are average values per scene.}
    \label{tab:ec_ablation}
    \begin{tabular}{rccc}
        \toprule
        \textbf{Error Correction} & \textit{\textbf{Pref rate}} & \textit{\textbf{\# Errors} }& \textit{\textbf{Run time}} \\
        \midrule
        No solver                       & 60.0\% & 12.3 & 0s \\
        Gradient descent                & 61.4\% & 0.5 & 2.6s \\
        LLM self-repair                 & 68.1\% & 5.7 & 106.7s \\
        Local search, basic imperative  & 71.4\% & 0.8 & 25.2s \\
        Local search, PSDL (ours)       & - & 1.1 & 9.3s \\
        \bottomrule
    \end{tabular}
\end{table}

\section{Results and Evaluation}
\label{sec:results}
In this section, we evaluate different layout generation approaches on their ability to synthesize open-universe 3D scenes. We compare our method with two declarative baselines and assess the impact of error correction through forced-choice perceptual studies. We also compare various ablation conditions using a new automated evaluation method, which we show is better aligned with the perceptual study results than previous methods for automatic evaluation of text-based visual generation systems.

\paragraph{Implementation Details}
Unless otherwise specified, we use Anthropic's \texttt{claude-3-5-sonnet-20241022} for language generation components of our proposed system.
We use OpenAI's \texttt{gpt-4o} as the multimodal LLM backbone for  automated evaluation methods.

\paragraph{Benchmark}
To evaluate layout generation methods, we created a benchmark of 70 scene prompts spanning diverse environments and complexity levels. Each prompt is labeled by four attributes: \emph{size} (small, medium, large), \emph{location} (indoor, outdoor), \emph{realism} (realistic, fantastical), and \emph{structure} (chaotic, structured).
The benchmark includes 14 small, 35 medium, and 21 large scenes; 48 indoor and 22 outdoor; 50 realistic and 20 fantastical; and 31 chaotic and 39 structured.

We also incorporate 52 prompts from the MIT indoor scenes dataset~\cite{Quattoni2009IndoorScenes} used by Holodeck in their evaluations. To include longer prompts, we also included 14 prompts that Holodeck uses for their qualitative results.
From our 70 prompts and 66 prompts from Holodeck, we extracted 19 long prompts we deemed as ‘complex’ to test more detailed prompts.
See the supplemental for the complete list of scene prompts.

\paragraph{Comparison Conditions}
In the subsequent experiments, we compare the following LLM-based scene layout generation methods:
\begin{itemize}
    \item \textbf{Ours}: generating layouts as procedural programs and then improving them using our iterative error correction scheme.
    \item \textbf{\declarativebaseline}: declarative layout generation approach described in~\cite{aguinakang2024openuniverseindoorscenegeneration}.
    \item \textbf{Holodeck}: the ``Constraint-based Layout Design Module''
    % \footnote{See page 5 of the Holodeck paper.}
    of the Holodeck system~\cite{yang2023holodeck}.
    \item \textbf{FlairGPT}: declarative layout generation approach described in~\cite{Littlefair2025_FlairGPT}.
\end{itemize}
We do not compare to LayoutGPT, as it is strictly dominated by Holodeck and \declarativebaseline, making this comparison unlikely to offer new insights.

\subsection{Human Perceptual Evaluation}
\label{sec:perceptual_study}
To compare how well different layout generation methods can produce layouts satisfying the scene prompts in our benchmark, we conducted two-alternative, forced-choice perceptual studies pitting our method against two of the declarative methods.

We recruited 10 participants from a population of university students.
%\edit{The set of participants does not intersect with the set of authors of the paper. Participants were not paid.}
The participants were divided into two groups, one for each study. Each participant was shown a series of 70 comparisons, where each comparison contains a scene prompt, images of two layouts in randomized order, and a question asking them to choose which scene they thought was better (taking into account overall scene plausibility and appropriateness for the prompt).
%\edit{We did not measure the time it took participants to perform the study, but the time was not limited. No data was excluded.}

For each comparison, we take the majority vote across all participants as the final answer.
Since we seek to evaluate only the quality of object layouts, to eliminate any impact that 3D model choice might have on participant response, objects in images were rendered as colored boxes over which participants could hover their mouse cursor to reveal the object's name.

Table~\ref{tab:perceptualstudy} shows the results of this experiment, and Figure~\ref{fig:qual_compare} shows some qualitative comparisons between generated layouts. Overall, participants preferred layouts generated using our method to those generated by either of the declarative versions. Examining individual decisions rather than majority vote, the per-task histograms (Supplementary Fig.~\ref{fig:preference-hist}, top) are strongly skewed toward our method: against DeclBase most tasks fall in the $4/5$--$5/5$ bins, and against Holodeck fully $51/70$ tasks are unanimous ($5/5$) for ours. The per-participant totals (Supplementary Fig.~\ref{fig:preference-hist}, bottom) show the same trend: each rater preferred our method on $48$--$61$ of $70$ scenes vs. DeclBase and on $61$--$68$ of $70$ vs. Holodeck.

%\edit{Looking at the individual decisions instead of the majority vote, we sorted tasks into bins where each bin corresponds to $0/5$, $1/5$, $2/5$, $3/5$, $4/5$ and $5/5$ participant votes for our method. We found that participants have a strong preference for our method against DeclBase ($2$, $4$, $6$, $15$, $20$, $23$ in bins), and especially against Holodeck ($0$, $1$, $3$, $4$, $11$, $51$ in bins). Looking at the preference breakdown by participant, individual participants preferred $48$, $50$, $48$, $61$, $49$ scenes generated by our method out of 70 against DeclBase and $64$, $68$, $63$, $61$, $62$ against Holodeck.}

To evaluate the impact of error correction, we conducted two other perceptual studies using the same protocol. Our method with error correction was preferred over our method without correction 74.3\% of the time, which in turn was preferred over DeclBase 61.8\% of the time. These results confirm that while describing scenes as procedural programs alone is effective, error correction plays a crucial role in further improving scene quality.

%As Holodeck is so strongly dis-preferred overall, there are not obvious trends in how scene types correlate with preference.
%In the comparison against \declarativebaseline, there is a bigger preference gap for larger and less chaotic scenes.
%These results suggest that while the imperative approach is effective overall, it is especially well-suited for large, dense scenes with considerable structure in their layouts.

\subsection{Automated Evaluation}
As perceptual studies are costly to run, we also investigated using automated evaluation metrics to approximate the results of a perceptual study.
We experimented with two metrics designed for automated evaluation of text-to-image generative models:
\begin{itemize}
    \item \textbf{VQAScore}~\cite{lin2024evaluating}: Scores how well a generated image matches a text prompt using the probability a visual question answering model assigns to the output token `yes' when asked whether the image depicts that text prompt. 
    \item \textbf{Davidsonian Scene Graphs (DSG)}~\cite{JaeminCho2024}: Computes a score by generating a dependency graph of simpler yes/no questions to ask about the image and then aggregating the percentage of those questions for which a VQA system returns 'yes.'
\end{itemize}
We can use these methods to compare two scene layouts by running them on rendered images of both and returning whichever has the higher score.
Unfortunately, we found that neither of these methods performed much better than chance (50\%) at agreeing with the majority-vote judgments from our perceptual study.
As long as a scene contains the right objects, VQA models can recognize the type of scene even with poor layouts, making VQAScore insufficiently discriminative for layout evaluation. Similarly, DSG struggled to generate yes/no questions which could differentiate the two scenes, leading to ties for most judgments.

These results motivated us to develop a simple new method for automated evaluation of scene layouts.
Specifically, we prompt a multimodal LLM with images of two scene layouts (along with a general task prompt) and ask it to list the pros and cons of each layout with respect to the scene prompt.
At the end of its output, the LLM returns which scene is better.
As shown in Table~\ref{tab:humanalign}, this simple method is much more aligned with human judgments.
Table~\ref{tab:humanalign} also includes results for an ablated version of our method which does not ask the LLM to first generate a pros \& cons list, illustrating that this additional step does improve the method's agreement with people.

As shown in Table~\ref{tab:scene_comp}, running our automated evaluation metric on our benchmark results in our scenes being chosen 77.1\% of the time over \declarativebaseline~scenes (vs. 82.9\% in the perceptual study) and 90.0\% of the time over Holodeck scenes (vs. 94.3\% from the perceptual study). While there is some discrepancy from `gold standard' human judgments, the trends are still clear.
%As shown in Table~\ref{tab:scene_comp}, running our automated evaluation metric on our benchmark and Holodeck's benchmark results in our scenes being chosen 76.5\% of the time over \declarativebaseline~scenes, 82.4\% of the time over Holodeck scenes, and 88.2\% of the time over FlairGPT scenes. Our scenes outperform all baseline methods across both benchmark prompt sets. Additionally, they demonstrate similar — and at times superior — performance with longer prompts.

Table~\ref{tab:complex_comp} shows that our automated evaluation metric prefers our scenes over Holodeck scenes across different object count scenarios. It is notable that our scenes are preferred the most when there are 40 or more objects, showing that our method excels in complicated scenes with high object counts.

\subsection{Error Correction}
%Table~\ref{tab:llm_results_restructured} reports how many imperative layout errors our error correction scheme fixes across our scene prompt benchmark using different LLM backbones.
%Different LLMs may produce layout programs that are more or less suitable for error correction, based on how they parameterize the layout.
%Claude 3.5 Sonnet's layouts are most amenable to correction, resulting in the fewest overall errors after correction is applied.
%Interestingly, OpenAI's inference-time compute model o1 is not noticeably better than GPT-4o in this case.

Table~\ref{tab:ec_ablation} reports automatic evaluation preference rates among various error correction strategies, and the average amounts of remaining errors. We can see that our local search for PSDL programs is very effective, correcting 91\% of all errors. Gradient descent and local search for basic imperative programs corrects even a bigger share of errors (96\% and 93.5\% respectively), because these two methods are able to move individual objects. However, our method wins over all baselines in terms of preference rates: as we explain in Section~\ref{sec:psdl}, resolving errors in the space of individual objects often breaks intended relationships between objects.

We describe the details of our implementation of LLM self-repair in the supplementary.

% Table~\ref{tab:llm_results} reports how effectively our correction procedure reduces layout errors from state-of-the-art LLM outputs. We observe that modern LLMs still struggle with numeric reasoning, leading to overlaps and out-of-bounds placements. Notably, the imperative paradigm does not inherently produce fewer errors than a coordinate-based approach; rather, the key advantage is that our error-correction mechanism can systematically fix these errors without disrupting the overall scene structure. Numerically, our error correction procedure brings the amount of errors from $17.65$ to $3.46$ on average. Finally, inference-time compute models (e.g., OpenAI's o1) offer little additional benefit in our experiments.

\subsection{Timing}
The scene template generation stage of our pipeline takes 9.5s, generating a PSDL layout program takes 19.2s, and error correction via program search takes 9.3s, on average. Overall, it takes 38s on average to synthesize a scene using our method. In comparison, synthesizing a \declarativebaseline~scene takes 40.8s on average, which consists of 10s for synthesizing a declarative layout program, 21.3s for the solver, and shares the template generation stage. Therefore, the average run time of our system is comparable to declarative systems.

Table~\ref{tab:ec_ablation} reports timings for various error correction schemes: gradient descent is the fastest, while LLM self-repair is more than an order of magnitude slower than our method. Program search for PSDL programs is nearly 3 times faster than for the basic imperative programs because PSDL's loops and variables lower the dimensionality of the search space.
\section{Conclusion and Future Work}
\label{sec:conclusion}
We introduced Procedural Scene Description Language (PSDL) and a complementary, LLM-free error-correction strategy that together re-establish the strengths of imperative scene layout programs while overcoming their fragility to LLM errors. By embedding explicit geometric relationships, shared variables, and structured control flow directly into a lightweight Python DSL, our approach enables fast, solver-free execution and yields a compact, semantically meaningful search space for symbolic repair. Across a diverse benchmark of open-universe prompts, perceptual studies and a new VLM-based automatic metric show that PSDL with program search delivers layouts that people prefer to those produced by state-of-the-art declarative and imperative baselines --- especially in large, highly structured scenes --- without incurring additional LLM calls.

%We introduced a new method for open-universe scene layout generation that adopts an imperative paradigm, in contrast to the declarative approaches commonly used in prior work.
%We proposed an iterative, LLM-free error correction mechanism, which refines generated scenes by adjusting scene parameters to improve validity while remaining close to the original layout.
%We evaluated our approach through a forced-choice perceptual study, showing that participants preferred scenes generated by our method over two declarative baselines 82\% and 94\% of the time, respectively. Finally, we also introduced a novel automated evaluation metric for judging scene layouts, demonstrating that (1) scenes generated by our method achieve higher scores compared to alternatives, and (2) this metric aligns more closely with human preferences than existing automated evaluation metrics.

\vspace{-0.7em}
\paragraph{Limitations}  
First, our local-search repair currently edits only numeric constants and facing directions; more complex semantic flaws (e.g., swapping object identities or swapping $x$ and $y$ coordinates) remain out of scope. Second, the loss we minimize does not explicitly encode higher-order aesthetics such as line-of-sight, walkability or affordances beyond support and collision. Third, while the VLM-based evaluator aligns better with human judgments than prior automatic scores, it is itself an empirical proxy whose biases may steer future systems toward its own blind spots. Finally, our experiments rely on a fixed object retrieval module and constrain orientations to four cardinals; relaxing either assumption will likely expose new challenges for both generation and correction.
%Despite its advantages, our method has certain limitations. First, object orientation in our pipeline is restricted to four cardinal directions. Extending this to support arbitrary orientations is necessary for handling a wider variety of scenes. 
%Further, the error correction mechanism, although more efficient than self-repair methods, is not fully inexpensive and is limited to parametric adjustments. It cannot address errors stemming from incorrect object retrieval or misinterpretation of input text prompts. Addressing such errors may require a hybrid approach that integrates symbolic error correction, like ours, with LLM-supported self-correction mechanisms, and warrants further investigation.
% Second, our pipeline depends on object retrieval, which can fail if the retrieval process is unreliable or the object dataset is insufficient. While this issue occurred rarely in our experiments, better integration of layout synthesis and object retrieval remains an open challenge.

\vspace{-0.7em}
\paragraph{Future Work}  
While scene generation has received significant attention in research, much of it has focused on static scenes. A natural next step is to extend scene synthesis to dynamic scenes, where objects interact or evolve over time. As the generation tasks grow in complexity, simplifying the coding process for LLMs, as done with our procedural approach, will become increasingly important.

\bibliographystyle{ACM-Reference-Format}
\bibliography{main}
\clearpage

\appendix
\section{Holodeck Modifications}
To focus on evaluating the quality of object layouts and eliminate any influence object selection might have on participant responses in the perceptual study, Holodeck’s object selection module was modified to use only the same set of objects and sizes present in the corresponding scenes from our system. To avoid prompting the LLM for the object selection plan json, which resulted in hallucinations of objects and object sizes outside of the given constraints, a 'mock' json following the LLM output format was manually created and inserted into the system pipeline for the layout module to use. Because Holodeck’s original pipeline handled both object selection and secondary object relations (e.g., placing monitors on desks) simultaneously, bypassing the LLM for object selection caused secondary relations to be omitted. As a result, Holodeck layouts for the 70 prompts used in the perceptual study lack secondary relations (e.g., computers appearing beside rather than on top of desks). We later use 
an updated configuration for the other set of 66 prompts, where secondary relations are included by prompting the LLM only for the object-to-object relations, without altering the fixed object list. Despite the two different configurations, the trend in preference rates of 'Ours' vs. 'Holodeck' in Table ~4 are evident, and we do not have reason to believe that they would drastically change if the 70 scenes were re-run with secondary relations. 

\section{LLM self-repair}
To implement LLM-based self-repair, we provide the language model with the original layout program and a description of the layout errors produced during execution. At each iteration, the LLM is prompted with a structured conversation: the system prompt describes the DSL and expected behaviors, the user provides the scene name, the assistant echoes the original layout code, and the user follows up with the list of detected errors (e.g., object overlaps, out-of-bounds issues, or support violations). The LLM then returns a revised layout program attempting to correct those issues.

This revised program is executed, and if errors remain, the process is repeated using the new scene as the next reference. We allow up to 6 correction iterations, as we observed that most successful repairs occur within the first few rounds and diminishing returns follow.

\begin{table}[t!]
    \renewcommand{\arraystretch}{0.8}
    \centering
    \caption{Preference rates for each scene category in the forced-choice perceptual study, comparing our method against \declarativebaseline~and Holodeck.}
    \label{tab:percep_study_details}
    \begin{tabular}{rcc}
        \toprule
        \textbf{Scene Type} & \textbf{\declarativebaseline} & \textbf{Holodeck}\\
        \midrule
        Ours w/ EC vs.   & 82.9\%  & 94.3\%\\
        \midrule
        Small            & 71.4\% & 100.0\%\\
        Medium           & 82.9\% & 91.4\%\\
        Large            & 90.5\% & 95.2\%\\
        \midrule
        Indoor           & 81.3\% & 95.8\%\\
        Outdoor          & 86.4\% & 90.9\%\\
        \midrule
        Realistic        & 84.0\%    & 92.0\%\\
        Fantastical      & 80.0\%    & 100.0\%\\
        \midrule
        Chaotic          & 74.2\% & 96.8\%\\
        Structured       & 89.7\% & 92.3\%\\
        \bottomrule
    \end{tabular}
\end{table}

\begin{table*}[t]
\small
\setlength{\tabcolsep}{5pt}
\renewcommand{\arraystretch}{1.12}
\centering
\caption{PSDL API (I): Types, attributes, and enumerations. Vector fields support component selection \texttt{.x}, \texttt{.y}, \texttt{.z}. Attributes marked R/W are writable via overloaded assignment; others are read-only inputs from the template.}
\label{tab:psdl-types}
\begin{tabularx}{\textwidth}{l l l c Y}
\toprule
\textbf{Symbol} & \textbf{Kind} & \textbf{Type} & \textbf{R/W} & \textbf{Meaning} \\
\midrule
\texttt{scene}      & special object & Object     & R   & Bounding cuboid of the whole scene; exposes \texttt{center}, \texttt{min}, \texttt{max}, \texttt{width}, \texttt{depth}, \texttt{height}. \\
\texttt{o}          & scene object   & Object     & R   & Any pre-instantiated object bound by the template (e.g., \texttt{chair}, \texttt{table}). \\
\texttt{o.center}   & attribute      & \texttt{vec3} & R/W & Geometric center of \texttt{o} in the current coordinate frame. Writing \texttt{o.center.\{x|y|z\}} moves \texttt{o} along that axis. \\
\texttt{o.min}      & attribute      & \texttt{vec3} & R/W & AABB minimum in the current frame; writing components (e.g., \texttt{o.min.x = \dots}) translates \texttt{o} to satisfy the constraint. \\
\texttt{o.max}      & attribute      & \texttt{vec3} & R/W & AABB maximum in the current frame. \\
\texttt{o.width}    & attribute      & \texttt{float} & R  & Object width (perpendicular to facing). \\
\texttt{o.depth}    & attribute      & \texttt{float} & R  & Object depth (along facing). \\
\texttt{o.height}   & attribute      & \texttt{float} & R  & Object height (upwards). \\
\texttt{o.facing}   & attribute      & \texttt{Facing} & R/W &
Orientation of \texttt{o} relative to the current frame. Allowed assignments:
\texttt{o.facing = dir} (set cardinal), \texttt{o.facing = p} (face \emph{toward} object \texttt{p}; forward axis points from \texttt{o.center} to \texttt{p.center} and is quantized to the nearest cardinal), or \texttt{o.facing = p.facing} (copy \texttt{p}'s orientation). The stored value is always a cardinal. \\
\texttt{o.name}     & attribute      & \texttt{str}   & R  & Object identifier from the template. \\
\texttt{o.support}  & attribute      & \texttt{Support} & R & Physical support type from the template (\texttt{STANDING}, \texttt{WALL\_MOUNTED}, \texttt{FLOATING}); not modifiable in PSDL. \\
\midrule
\texttt{Facing}     & enum           & ---        & --- & Cardinal orientation: \texttt{X\_NEG}, \texttt{X\_POS}, \texttt{Y\_NEG}, \texttt{Y\_POS}. (Assignment to \texttt{o.facing} is overloaded as described above; the stored value remains a cardinal.) \\
\texttt{Support}    & enum           & ---        & --- & \texttt{STANDING}, \texttt{WALL\_MOUNTED}, \texttt{FLOATING}. \\
\bottomrule
\end{tabularx}
\end{table*}

\begin{table*}[t]
\small
\setlength{\tabcolsep}{6pt}
\renewcommand{\arraystretch}{1.12}
\centering
\caption{PSDL API (II): Statements/operators and assignment forms. Right-hand sides (RHS) are standard Python expressions composed of numbers, Python variables, arithmetic, and attribute reads (e.g., \texttt{p.center.x}, \texttt{q.min.z}, \texttt{scene.max.y}).}
\label{tab:psdl-ops}
\begin{tabularx}{\textwidth}{l l Y}
\toprule
\textbf{Construct} & \textbf{Form / Signature} & \textbf{Semantics and Examples} \\
\midrule
Local frame & \texttt{set\_coordinate\_frame(o)} &
Set the current coordinate frame to object \texttt{o}: y-axis aligns with \texttt{o.facing}, x-axis is $90^\circ$ clockwise from y, z-axis is up. Useful for building sub-layouts in canonical space. Example: \texttt{set\_coordinate\_frame(counter)}. \\
\midrule
Center component assignment & \texttt{o.center.\{x|y|z\} = expr} &
Moves \texttt{o} so the specified center component equals \texttt{expr} (in the current frame). Example: \texttt{chair.center.y = table.center.y}. \\
AABB component assignment & \texttt{o.min.\{x|y|z\} = expr} / \texttt{o.max.\{x|y|z\} = expr} &
Translates \texttt{o} so the chosen AABB face matches \texttt{expr}. Example: \texttt{chair.max.x = table.min.x - 0.1}. \\
Facing assignment (cardinal) & \texttt{o.facing = dir} &
Sets \texttt{o}’s orientation to a cardinal \texttt{dir}. Example: \texttt{register.facing = X\_NEG}. \\
Facing assignment (face toward object) & \texttt{o.facing = p} &
Rotates \texttt{o} to face object \texttt{p}; the forward axis points from \texttt{o.center} to \texttt{p.center}, and the resulting \texttt{o.facing} is quantized to the nearest cardinal. Example: \texttt{chair.facing = table}. \\
Facing assignment (copy orientation) & \texttt{o.facing = p.facing} &
Copies the orientation of \texttt{p}. Example: \texttt{chair.facing = table.facing}. \\
\midrule
RHS expressions & (Python) &
Any Python arithmetic expression over scalars/variables and attribute reads from \texttt{scene} or previously placed objects. Examples: \texttt{scene.center.x + i * d}, \texttt{counter.min.x + (i+0.5) * counter.width}. \\
Control flow & (Python) &
Ordinary Python loops and conditionals (\texttt{for}, \texttt{if}, \texttt{enumerate}, \texttt{range}, \dots) to express repetition and symmetry. Example: \texttt{for i, stool in enumerate(stools):\quad stool.center.x = counter.min.x + (i+1.0) / 4.0 * counter.width}. \\
\bottomrule
\end{tabularx}
\end{table*}

\begin{table*}[t]
\centering
\small
\caption{Comparison of coordination errors, out-of-bounds placements, and overlaps for different LLM configurations across three setups: explicit coordinate prediction, pre-correction, and post-correction.}
\label{tab:llm_results_restructured}
\begin{tabular}{cccc|ccc|ccc|ccc}
\toprule
\multirow{2}{*}{\centering\textbf{Model}}& 
\multicolumn{3}{c}{\texttt{\small claude-3-5-sonnet-20241022}} &
\multicolumn{3}{c}{\texttt{\small gpt-4o-2024-11-20}} &
\multicolumn{3}{c}{\texttt{\small o1-2024-12-17}} &
\multicolumn{3}{c}{\texttt{\small gemini-exp-1206}} \\
\cmidrule(lr){2-13}
 & \textsc{All} & \textsc{Bound} & \textsc{Ovl} 
 & \textsc{All} & \textsc{Bound} & \textsc{Ovl} 
 & \textsc{All} & \textsc{Bound} & \textsc{Ovl} 
 & \textsc{All} & \textsc{Bound} & \textsc{Ovl} \\
\hline
% \textbf{\small Explicit Coordinates} & 19.50 & 2.00 & 11.29 & 21.79 & 3.30 & 10.10 & 17.20 & 2.36 & 10.63 & 13.07 & 1.70 &  7.66 \\
\textbf{\small Ours w/o correction}  & 17.57 & 1.56 & 10.76 & 17.80 & 1.14 & 11.19 & 17.29 & 2.36 & 10.70 & 14.67 & 1.19 & 10.41 \\
\textbf{\small Ours} & \textbf{2.10}  & \textbf{0.54} & \textbf{0.56}  & \textbf{3.34}  & \textbf{0.61} & \textbf{0.64}  & \textbf{3.24}  & \textbf{0.63} & \textbf{1.17}  & \textbf{3.14} & \textbf{0.61} & \textbf{1.14} \\
\bottomrule
\end{tabular}
\end{table*}

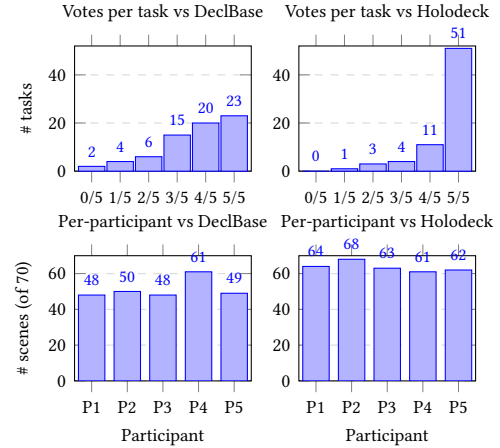
\begin{figure}[t]
\centering
\begin{tikzpicture}
\begin{groupplot}[
  group style={group size=2 by 2, horizontal sep=18pt, vertical sep=32pt},
  width=0.46\linewidth, height=0.38\linewidth,
  %ybar, bar width=10pt,
  every axis/.append style={
    ybar, bar width=10pt,
  },
  ymajorgrids=true, grid style={dashed,gray!30},
  nodes near coords,
  nodes near coords align={vertical},
  nodes near coords style={font=\scriptsize,anchor=south,yshift=0.5pt},
  tick label style={font=\footnotesize},
  label style={font=\footnotesize},
  title style={font=\footnotesize},
  enlarge x limits=0.12
]

\nextgroupplot[
  title={Votes per task vs DeclBase},
  ylabel={\# tasks},
  ymin=0, ymax=52,
  xtick=data,
  symbolic x coords={0/5,1/5,2/5,3/5,4/5,5/5}
]
\addplot coordinates {(0/5,2) (1/5,4) (2/5,6) (3/5,15) (4/5,20) (5/5,23)};

\nextgroupplot[
  title={Votes per task vs Holodeck},
  ymin=0, ymax=52,
  xtick=data,
  symbolic x coords={0/5,1/5,2/5,3/5,4/5,5/5}
]
\addplot coordinates {(0/5,0) (1/5,1) (2/5,3) (3/5,4) (4/5,11) (5/5,51)};

\nextgroupplot[
  title={Per-participant vs DeclBase},
  ylabel={\# scenes (of 70)},
  xlabel={Participant},
  ymin=0, ymax=70,
  xtick=data,
  symbolic x coords={P1,P2,P3,P4,P5}
]
\addplot coordinates {(P1,48) (P2,50) (P3,48) (P4,61) (P5,49)};

\nextgroupplot[
  title={Per-participant vs Holodeck},
  xlabel={Participant},
  ymin=0, ymax=70,
  xtick=data,
  symbolic x coords={P1,P2,P3,P4,P5}
]
\addplot coordinates {(P1,64) (P2,68) (P3,63) (P4,61) (P5,62)};

\end{groupplot}
\end{tikzpicture}
\caption{\editG{Human preference study. Top row: per-task vote counts for our method (0/5–5/5) vs DeclBase (left) and Holodeck (right); each sums to 70 tasks. Bottom row: per-participant counts (out of 70) preferred for our method against each baseline.}}
\label{fig:preference-hist}
\end{figure} %!!!

\begin{figure*}[ht!]
    \centering
    \includegraphics[width=0.9\linewidth]{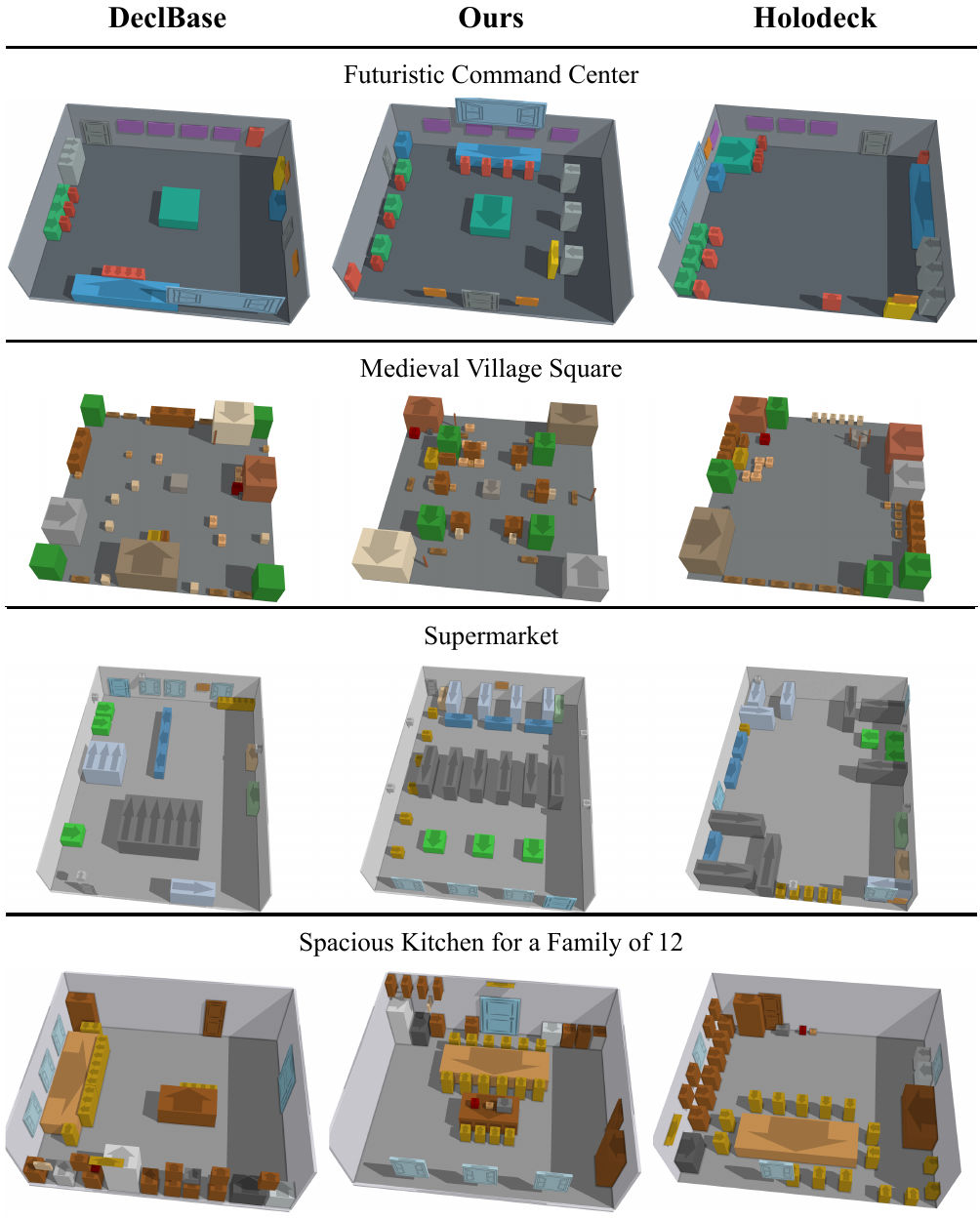}
    \caption{
    Qualitative comparisons between our method, DeclBase, and Holodeck. Our method and Holodeck uses \texttt{gpt-4o}, while DeclBase uses \texttt{claude-3-5-sonnet-20241022}. See the supplemental for a comparison between our method and DeclBase only using \texttt{claude-3-5-sonnet-20241022}.
    }
    \label{fig:qual_compare}
\end{figure*}

\begin{figure*}[ht!]
    \centering
    \includegraphics[width=\linewidth]{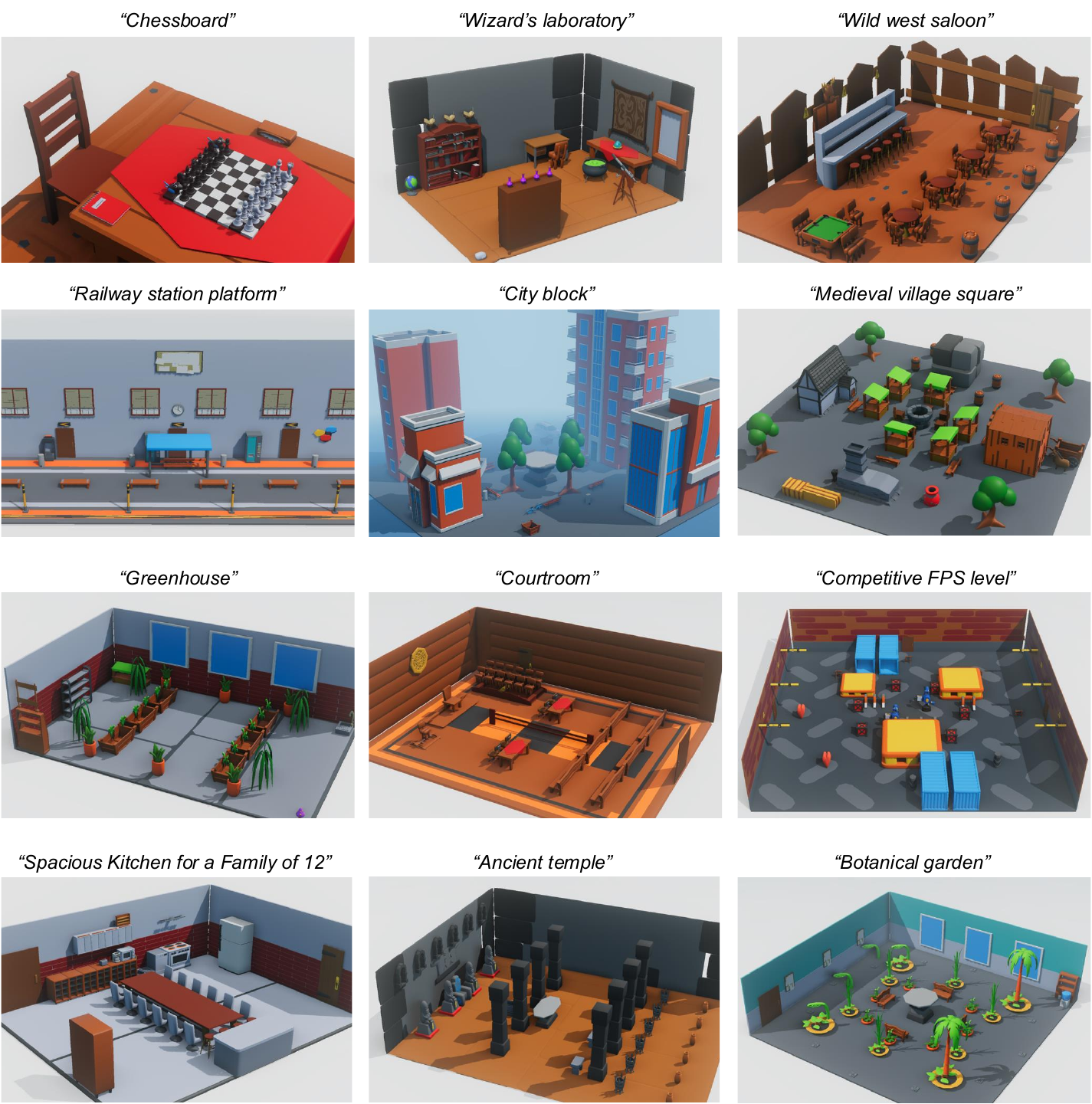}
    \caption{
    More scenes synthesized using our imperative layout generation method with error correction.
    }
    \label{fig:qualitative_hh}
\end{figure*}

\clearpage
\appendix

% \kenny {TODO REMOVE SUPP FROM SUBMISSION}
%\input{supp_content}

\end{document}